\documentclass[twocolumn]{svjour3}         
\smartqed  
\usepackage{graphicx}
\usepackage{natbib}
\usepackage{amssymb}

\newcommand{\apss}{Ap\&SS}
\newcommand{\apj}{ApJ}
\newcommand{\aj}{AJ}
\newcommand{\apjl}{ApJ}
\newcommand{\apjs}{ApJS}
\newcommand{\aap}{A\&A}
\newcommand{\mnras}{MNRAS}
\newcommand{\nat}{Nature}

\journalname{Astrophysics \& Space Science}
\begin{document}

\title{Star and Planet Formation with ALMA: an Overview
}

\titlerunning{Star and Planet Formation with ALMA}   

\author{Ewine F. van Dishoeck         \and
        Jes K.\ J{\o}rgensen 
}


\institute{Leiden Observatory \at
              P.O. Box 9513, 2300 RA Leiden, The Netherlands \\
              \email{ewine@strw.leidenuniv.nl}           
           \and
           Harvard-Smithsonian Center for Astrophysics \at
           60 Garden Street MS~42, Cambridge, MA 02138
}

\date{Received: May 13, 2007 / Accepted: July 13, 2007}

\maketitle

\begin{abstract}
  Submillimeter observations with ALMA will be the essential next step
  in our understanding of how stars and planets form. Key projects
  range from detailed imaging of the collapse of pre-stellar cores and
  measuring the accretion rate of matter onto deeply embedded
  protostars, to unravelling the chemistry and dynamics of high-mass
  star-forming clusters and high-spatial resolution studies of
  protoplanetary disks down to the 1~AU scale.

\end{abstract}

\section{Introduction}

The formation of stars and planets occurs deep inside clouds and disks
of gas and dust with hundreds of magnitudes of extinction, and can
therefore only be studied at long wavelengths.  
In the standard scenario for the formation of an isolated low-mass
star, a cold core contracts as magnetic and turbulent support are lost
and subsequently collapses from the inside out to form a protostar
with a surrounding disk.  Soon after formation, a stellar wind breaks
out along the rotational axis of the system and drives a bipolar
outflow entraining surrounding cloud material.  The outflow gradually
disperses the protostellar envelope, revealing an optically visible
pre-main sequence star with a disk.  Inside this disk, grains collide
and stick owing to the high densities, leading to pebbles, rocks and
eventually planetesimals which interact to form planets. The original
interstellar gas and dust is gradually lost from the disk through a
combination of processes, including accretion onto the new star,
formation of gas-rich planets, photoevaporation and stellar
winds. 

These different evolutionary stages in star- and planet formation are
traditionally linked to their Spectral Energy Distributions (SEDs)
\citep{Lada99}, which illustrate how the bulk of the luminosity shifts
from far- to near-infrared wavelengths as matter moves from envelope
to disk to star. So far, most tests of this scenario have been done
using spatially unresolved data which encompass the entire
star-disk-envelope system in a single beam. ALMA will be the first
telescope capable of spatially and spectrally resolving the individual
components and tracing the key physical and chemical processes on all scales.

The strengths of ALMA are (i) its high angular resolution, combined
with enough sensitivity to image continuum and lines down to 0.01$''$
($\sim 1$~AU = terrestrial planet-forming zone at 150 pc, 30 AU = disk
of high-mass YSO at 3 kpc); (ii) its high spectral resolution down to
0.01 km s$^{-1}$ so that the details of the dynamics and kinematics
can be probed; (iii) access to thousands of lines from hundreds of
species allowing a wide variety of physical and chemical regimes to be
probed; and (iv) its ability to detect optically thin dust emission
and thus directly derive dust masses.

ALMA probes the wavelength range of 0.3--9 millimeter, which is on the
Rayleigh-Jeans tail of the SEDs of young stellar objects. For a
complete picture of these sources complementary space and ground-based
observations at shorter wavelengths are necessary. Compared with other
missions, ALMA is less well suited for large area surveys because of
its small field of view. Mid-infrared observatories such as the {\it
Spitzer Space Telescope} at 3--70 $\mu$m probe the peak of the SED for
low-mass YSOs and can scan large areas much more rapidly, albeit at
lower spatial and spectral resolutions.  Near-infrared imaging is a
powerful tool to characterize the stellar component, whereas the {\it
Herschel Space Observatory} and ground-based single-dish submillimeter
telescopes equipped with large format bolometers can rapidly search
large areas for cold dust emission.  These missions will provide
complete unbiased catalogs with thousands of sources covering all the
nearby molecular clouds and star-forming regions within a few hundred
pc (contained within Gould's Belt), many of the young clusters and
associations within 1 kpc, and most of the prominent high-mass
star-forming clouds to the outer edge of the Galaxy. Thus, the primary
source lists for ALMA will come from these missions. They will also
provide the main statistical results from which, for example,
timescales for the different phases can be derived.

This paper outlines a number of key questions for each evolutionary
state where ALMA can make a major contribution. It focusses mostly on
low-mass star formation and protoplanetary disks, but many of the same
arguments are also valid for high-mass star formation.
Inspiration for this review was provided by the many beautiful
paintings by Juan Mir\'o displayed in Madrid and elsewhere around the
world. Most appropriate for this topic are `Birth of the world',
'Chiffres and constellations', `Red disk' and `Serpent looking at
comet'. A challenge for the reader is to find the relations
between these paintings and the topics described here.

\section{Low-mass star formation}

\subsection{Pre-stellar cores}\label{prestellar}

{\it Q1: What are the initial conditions for low-mass star formation,
in particular the physical structure and kinematics of the densest
part of the core?}

In recent years, a number of cold, highly extincted clouds have been
identified which have a clear central density condensation. These
so-called pre-stellar cores are believed to be on the verge of
collapse and thus represent the earliest stage in the star-formation
process \citep[e.g.,][]{tafalla98}. The physical and chemical state of
these clouds is now well established on scales of few thousand AU by
single dish millimeter observations combined with extinction maps. The
cores are cold, with temperatures varying from 10--15 K at the edge to
as low as 7--8 K at the center, and have density profiles that are
well described by Bonnor-Ebert profiles. It is now widely accepted
that most molecules are highly depleted in the inner denser parts of
these cores \citep{caselli99,bergin02}: images of clouds such as B68
show only a ring of C$^{18}$O emission, with more than 90\% frozen out
toward its center.

ALMA will be particularly powerful in probing the central part of the
core on scales of 100 AU and search for signs of collapse in the very
earliest stages. Important probes are the lines of N$_2$H$^+$ and
H$_2$D$^+$ at 372 GHz, with the latter line a unique probe of the
kinematics in regions where all heavy molecules are depleted
\citep{vandertak05}.

\subsection{Very low luminosity objects and formation of brown dwarfs}

{\it Q2: What prevents some clouds from collapsing? Why do some
low-luminosity sources have such a low accretion rate in spite of the
much larger reservoir of gas and dust?}

\begin{figure}
\resizebox{\hsize}{!}{\includegraphics{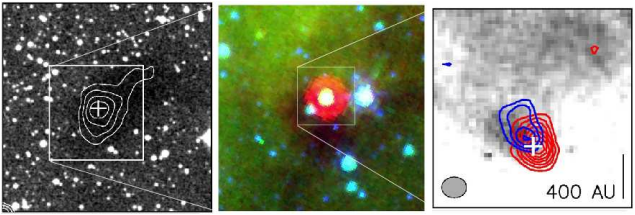}}
\caption{The VeLLO L1014-IRS. Left: optical image with 1.2~mm dust
  continuum emission overlayed. Middle: Spitzer mid-infrared image
  with 4.5~$\mu$m (blue), 8.0~$\mu$m (green) and 24~$\mu$m
  (red). Right: CO 2--1 map of the innermost region of the core from
  the SubMillimeter Array (SMA). Images from \cite{young04} (left, middle) and
  \cite{bourke05} (right).}\label{l1014fig}
\end{figure}

About 75\% of so-called `starless' cores (i.e., dark cores with no
IRAS source) remain starless down to 0.01 L$_{\odot}$ or less even
after deep surveys with {\it Spitzer} \citep{kirk07}.  However, {\it
Spitzer} has revealed a small set of cores with so-called Very Low
Luminosity Objects (VeLLOs). Examples include L1014 ($\sim$0.1
L$_{\odot}$, \citealt{young04}), L1521F ($\sim 0.05$ L$_{\odot}$
\citealt{crapsi05,bourke06}), and IRAM~04191 ($\sim 0.08$ L$_{\odot}$,
\citealt{dunham06}). These VeLLOs are embedded in cores with typical
masses of 1 M$_{\odot}$, but their low luminosities suggest that their
central stellar masses are low and that they (currently) have low
accretion rates. They also show very different outflow properties
ranging from a large well developed outflow in IRAM~04191 to a
miniscule outflow in L1014 only detectable through high angular
resolution millimeter observations (Fig.~\ref{l1014fig}). This
suggests that accretion in these cores may be episodic. It remains an
interesting question whether these VeLLOs constitute a separate stage
in the evolution of low-mass protostars or are precursors of
substellar objects, but without a better handle on their dynamical
structure it is difficult to predict the ``end result'' of the ongoing
star formation in these cores. ALMA will be able to zoom in on these
sources, image their disks (whose presence is inferred from the SEDs)
and small scale outflows and furthermore constrain the kinematics
of their envelopes.

{\it Q3: How do brown dwarfs form? Like stars or like planets?}

The VeLLOs described above could be the precursors of substellar
objects like brown dwarfs. If so, the presence of outflows and
accretion disks imply a formation process similar to that of
stars. Formation in, or fragmentation of, a disk around a more massive
primary star is not consistent with these data. Many other lines of
evidence based on infrared imaging and spectroscopy point in the same
direction \citep[e.g.,][]{testi02,mohanty04}.  Large samples of young
brown dwarfs now exist, several of them in wide binaries which are
easily disrupted, providing further clues on their origin \citep[see][for a review]{luhmanppv}. ALMA will be able to determine whether brown
dwarfs in the earliest stages have similar statistics in terms of
binary fraction and separation.

\subsection{Formation of stellar clusters and origin of the IMF}\label{imf}

{\it Q4: What is the relation between cloud or clump structure and the
IMF?}

The advent of large bolometer arrays on submillimeter telescopes has
revived detailed studies of the structure of molecular clouds and
cores just prior to and during the star formation process.  On large
pc-size scales, molecular clouds have a highly inhomogeneous or
`clumpy' structure, from which the mass distribution $\Delta N/\Delta
M$ can be measured (see \cite{williamsppiv} for a review).
Interestingly, the mass spectrum of cores follows a law $\propto$
$M^{-1.5}$ below 0.5 M$_{\odot}$ and $\propto$ $M^{-2.5}$ above 0.5
M$_{\odot}$ \citep{alves07}, similar to the stellar initial mass
function (IMF).  This suggests that the IMF may already be determined
at the pre-stellar stage during the fragmentation of a (turbulent)
molecular cloud, a result with wide-ranging implications for studying
the evolution of molecular clouds in our Galaxy and galaxies as a
whole.  The sensitivity and spatial resolution of ALMA are needed,
however, to separate the lower density cloud material from the dense
cores and to link this work with the optical and infrared
determinations of the low-mass end of the IMF in young clusters. For
example, the mass spectrum is still uncertain for clump masses below
0.1 M$_{\odot}$, whereas ALMA can probe the mass function down to
planetary masses and study the origin and distribution of brown dwarfs
and free-floating Jupiter-mass exo-planets using also kinematic
information.

{\it Q5: What fraction of stars forms as binaries or multiples? Are
all cluster members co-eval? Is there evidence for dynamical
interactions?}

Related to Q4 is the question whether cloud fragmentation leads to a
young cluster or to distributed star formation.  In contrast with
previous claims, {\it Spitzer} finds evidence for both processes,
including a significant fraction of young stars distributed throughout
the clouds \citep[e.g.,][]{allenppv}. The formation of the more
massive members of a cluster through competitive accretion is also
heavily debated \citep[e.g.,][]{bate05,krumholz05}. ALMA can address
this question by determining masses of YSOs in the earliest, deeply
embedded stages when accretion is still taking place. The fraction of
binaries and multiples can be compared with those of optically visible
T Tauri stars and field main-sequence stars, providing clues on binary
evolution and dynamical interactions (e.g., ejection).

{\it Spitzer} data show a wide variety of SEDs of clusters of YSOs on
0.1 pc scales \citep[e.g.,][]{rebull07,perspitz}. What causes this diversity?
The normal assumption is that the different SEDs (rising, falling)
reflect a real age spread from $<$0.1 to $>$1 Myr as part of a
more-or-less uniform star formation process. However, on scales as
small as 0.1 pc it is also reasonable to assume that all objects form
quasi-simultaneously from a fragmenting core. In such a co-eval
scenario, the diversity in SEDs would imply that objects go through
the evolutionary states at different rates.  ALMA and mid-infrared
data will be needed to settle this.

\subsection{Embedded YSOs: infall vs.\ outflow}

{\it Q6: What are the accretion rates during the earliest stages of
star formation and how do they vary with time?}

\begin{figure}
\resizebox{\hsize}{!}{\includegraphics[width=0.60\textwidth]{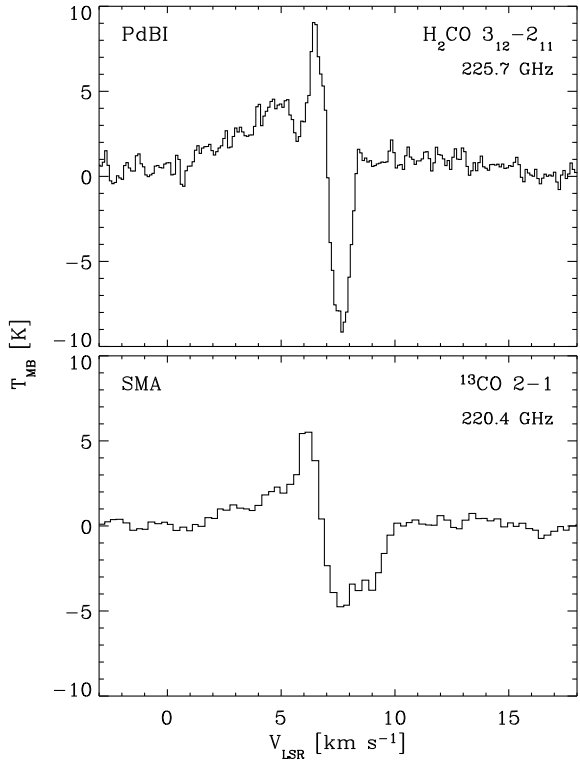}}
\caption{Inverse P~Cygni profiles toward the Class 0 YSO
NGC1333-IRAS4A in lines of H$_2$CO from IRAM Plateau de Bure
observations \citep{difrancesco01} and $^{13}$CO from SMA observations
\citep{prosacpaper}.}\label{invpcygni}
\end{figure}

Deeply embedded young stellar objects (the so-called `Class 0'
objects) have a complex physical and kinematical structure, with
envelopes, disks and outflows all blurred together in current
single-dish observations. High spatial and spectral resolution ALMA
data will be essential to disentangle the infall, outflow and rotation
components of these systems and study their evolution.
Of particular importance will be to measure directly the accretion
rates onto the disk and star. Redshifted absorption against the
continuum is thought to be the most direct tracer
(Fig.~\ref{invpcygni})  but this absorption is completely overwhelmed
by large-scale emission in single-dish data. Interferometer data
reveal such red-shifted absorption in a few (but not all) Class 0
sources but only in low-density tracers perhaps indicative of
large-scale infall rather than small-scale accretion
\citep[e.g.,][]{prosacpaper,difrancesco01}.  ALMA will have orders of
magnitude higher sensitivity to infall tracers due to the combination
of larger collecting area and smaller beam.

{\it Q7: What drives outflows and how does the outflow structure
change with time?}

\begin{figure}
\resizebox{\hsize}{!}{\includegraphics[angle=0]{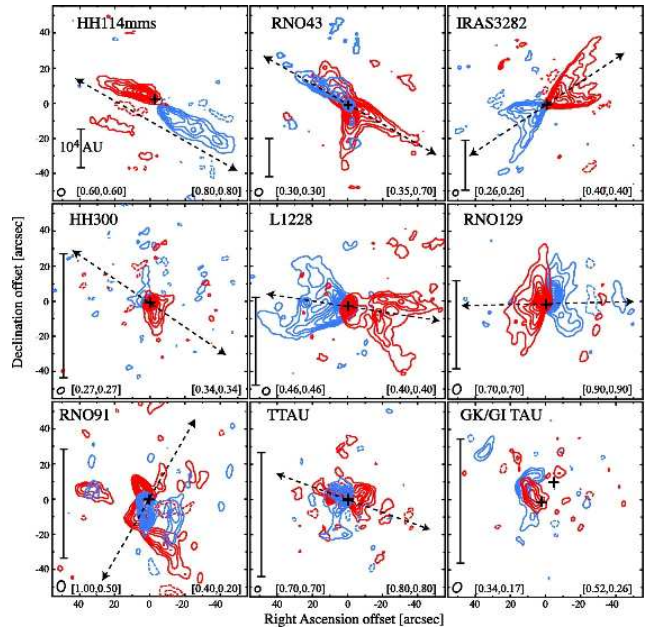}}
\caption{Gallery of $^{12}$CO 1--0 outflows from the OVRO survey of 
\cite{arce06} with Class 0 objects in the \emph{top}, Class I in the 
\emph{middle} and Class II in the \emph{bottom} panels. In each panel 
the cross mark the position of the protostar given by millimeter 
continuum observations.}\label{co_outflows}
\end{figure}

Violent outflows are a key characteristic of star formation
\citep{richerppiv}.  
Although they have been studied with single-dish telescopes for more
than 25 years, the mechanism of their formation remains poorly
understood \citep{shangppv}.  In the most deeply embedded objects,
highly collimated jet-like molecular outflows with extreme velocities
up to 200 km s$^{-1}$ are observed, but when the protostar evolves,
both the mechanical power and the collimation seem to decrease,
suggesting that the former is due to a decline in the overall mass
accretion rate \citep[e.g.,][]{bachillercreteII,arce06}
(Fig.~\ref{co_outflows}).  ALMA will permit studies of many different
kinds of YSOs, and, combined with independent estimates of the mass
accretion rate (see above), test magnetohydrodynamical models of their
evolution.  The actual location and mechanism by which outflows are
launched, and whether they are episodic in nature, is still a subject
of intense debate.  ALMA will provide detailed images of the
disk/outflow interface down to a few AU scales where the outflow is
accelerated and where the most intense interactions between the
outflow and its surroundings take place.

{\it Q8: What is the role of outflows in determining the final mass of
a star and in dispersing the core?}

Comparisons between cloud core and stellar mass functions show great
similarity in shape (see \S~\ref{imf}) but with a shift in mass of
typically a factor of $\sim$3 \citep{alves07}. Is this largely
due to the action of outflows dispersing the cores? Also, what is the
role of outflows in carrying off angular momentum? Attempts to measure
the mass dispersion rate in embedded YSOs have traditionally suffered
from poor spatial resolution.  Interferometer data are needed to
measure the local specific angular momentum as a function of radius in
the protostellar envelope, from the large-scale core to the
rotationally supported disk-like structures
\citep[e.g.,][]{hogerheijde01,takahashi06}.

\section{High-mass star formation}

{\it Q9. Is high-mass star formation a scaled-up version of low-mass star
formation? What triggers high-mass star formation?}

High-mass stars ($\gtrsim10$ M$_\odot$, $\gtrsim$10$^4$ L$_\odot$)
play a major role in the interstellar energy budget and the shaping of
the Galactic environment \citep[e.g.,][]{cesaroni05review}.  Phenomena
associated with massive stars such as photoionization, powerful winds,
shocks, expanding H II regions and supernovae drastically modify the
interstellar medium.  Due to large distances, short time scales, and
heavy extinction, the formation of high mass stars is still poorly
understood compared to that of their lower mass counterparts.  The
earliest stages of massive star formation have been revealed as dark
clouds seen in absorption against mid-infrared emission (the so-called
`Infrared Dark Clouds') \citep[e.g.,][]{simon06} and systematic
surveys are attempting to put the various observational signposts in
an evolutionary sequence starting with centrally condensed clouds with
masers (so-called High-Mass Protostellar Objects), followed by hot
cores with high temperature ($>100$ K) and high abundances complex
organic molecules (Hot Molecular Cores), and subsequently
ultra-compact H II regions showing significant amounts of ionized gas.

The similarity of some of these stages with those of their lower-mass
counterparts, coupled with the detection of outflows and signs of
rotating disks \citep[e.g.,][]{shepherd96,shepherd99}, suggest a
common formation mechanism for all stars. However, while these
similarities may hold for young B-type stars, the situation is much
less clear for O-type stars and the debate between the turbulent cloud
fragmentation scenario and that of competetive accretion or mergers is
still far from settled \citep{mckee03,krumholz05,bonnell06}.  A
related question is whether the formation of massive stars is largely
triggered, either by shocks compressing the cloud material (`collect
and collapse') or by `radiation-driven implosion'. Clear observational
evidence for triggered star formation is still lacking.
Examination of cluster properties and separating the more difffuse
clump and denser core material will be particularly relevant to
determine which process dominates in which environment.  Contrasting
star formation in the outer Galaxy, where metallicity, densities,
radiation field and gravitational potential well are lower compared
with the inner Galaxy may also be revealing.

\section{Astrochemistry}

{\it Q10. Which molecule is best suited to trace which physical
component of the YSO environment? Can we use chemistry as a `clock' of
the evolutonary state of the object?}

Systematic studies of molecules in YSOs are starting to reveal the
different chemical characteristics associated with the various stages
star formation \citep[e.g.,][]{paperii,ceccarellippv,vandishoeckpnas}.
The coldest pre-stellar cores show heavy freeze-out of virtually all
gas-phase molecules onto the cold grains (see \S\ref{prestellar}),
where grain-surface chemistry can lead to more complex species.  Once
the protostar starts to heat the envelope, the ices will evaporate in
a sequence according to their sublimation temperatures, with the most
volatile species like CO coming off at temperatures as low as 20 K and
the most strongly bound species like H$_2$O around 100 K.  The impact
of outflows on the inner envelope can also liberate molecules from the
ices and sputter grain cores. The evaporated ices subsequently drive a
rapid gas-phase chemistry for a period of $10^4-10^5$ yr. For example,
reactions with evaporated CH$_3$OH are thought to lead to high
abundances of CH$_3$OCH$_3$ and HCOOCH$_3$, although a grain surface
origin of these molecules is also possible
\citep[e.g.,][]{bisschop07,bottinelli07}.  These so-called `hot cores'
are signposts of the earliest stages of high-mass star formation, and
are now also found around some low-mass YSOs. After $\sim 10^5$ yr,
the abundances are reset by ion-molecule chemistry to their normal
cloud values, leading to the potential of molecules to act as chemical
clocks.

So far, this scenario is almost entirely based on spatially unresolved
single-dish data, with smaller-scale structure extracted from
observations of multiple lines of the same molecule with different
excitation conditions. Limited interferometer data confirm the
different chemical zones \citep[e.g.,][]{l483art}, but only ALMA will
have the combined sensitivity, spatial resolution and $(u,v)$ coverage
to make chemical `images' of YSOs in large sets of lines necessary to
directly test chemical models and explore and develop the use of
molecules as clocks of star formation.

{\it Q11. How far does chemical complexity go? Can we find (the
building blocks of) pre-biotic molecules?}

Of the $>$130 different molecules detected in interstellar clouds, the
majority ($\sim$75\%) are organics, including species as complex as
ethyl-cyanide (C$_2$H$_5$CN), acetamide (CH$_3$CONH$_2$, the largest
interstellar molecule with a peptide bond) and glycol-aldehyde
(CH$_2$OHCHO, the first interstellar sugar)
\citep[e.g.,][]{hollis00,hollis06}. However, in spite of literature
claims, the simplest amino-acid glycine (NH$_2$CH$_2$COOH) has not yet
been convincingly detected.  Current instrumentation prevents deep
searches for more complex molecules for several reasons: (a) the
regions of high chemical complexity are often very small ($<$1$''$),
the typical sizes of hot cores; (b) the crowding of lines is usually
so high that the confusion limit is reached; and (c) the largest
molecules have many close-lying energy levels so that the intensity is
spread over many different lines, each of them too weak to detect.
ALMA will be able to push the searches for prebiotic molecules two
orders of magnitude deeper to abundances of $< 10^{-13}$ with respect
to H$_2$, because it will have a much higher sensitivity to compact
emission and will resolve the sources so that spatial information can
be used to aid identifications of lines.

{\it Q12. Which fraction of complex molecules will end up unaltered in
the protoplanetary disk? How are they modified before incorporation
into planetary systems?}

The dynamics of gas in the inner few hundred AU of protostellar
envelopes are not yet well understood, but are important to determine
whether some of the observed (complex) molecules end up in the
rotating disk. Also, the disk entry point is highly relevant, since
gas falling in too close to the star will experience such a strong
accretion shock onto the disk that all molecules will dissociate
\citep{neufeld94} and only molecules entering at much larger distances
survive. ALMA's high spatial and kinematic resolution will obviously
be needed to address this issue.

Once in the disk, the chemistry is governed by similar gas-phase and
gas-grain interactions as in envelopes, but at higher densities. Also,
UV and X-rays from the young star dissociate molecules and modify the
chemistry in the optically thin surface layers.  This results in a
layered chemical structure, with a top layer consisting mostly of
atoms, a mid-plane layer where most molecules are frozen out, and an
intermediate layer where the dust grains are warm enough to prevent
complete freeze-out and where molecules are sufficiently shielded from
radiation to survive \citep[for a review, see][]{berginppv}. So far,
chemical images have been limited to just a few pixels across a
handful of disks in a few lines
\citep[e.g.,][]{qi03,pietu07}. Obviously, ALMA will throw this field
wide open. A particularly exciting topic is the chemistry in the inner
disk, i.e., inside the `snow-line' where all molecules evaporate and
the chemistry approaches that at LTE. For example, {\it Spitzer} data
have revealed highly abundant and hot HCN in the inner disk
\citep{lahuis06}. The brightness temperatures of the submillimeter
lines are predicted to be several hundred K, sufficient for ALMA to
image the inner few AU in the nearest disks.

\section{Protoplanetary disks}

\subsection{Young disks in the embedded phase}

{\it Q13. How do disks form and grow with time? How hot or cold is the
disk? Can we find evidence for gravitational instabilities?}

\begin{figure}
\resizebox{\hsize}{!}{\includegraphics{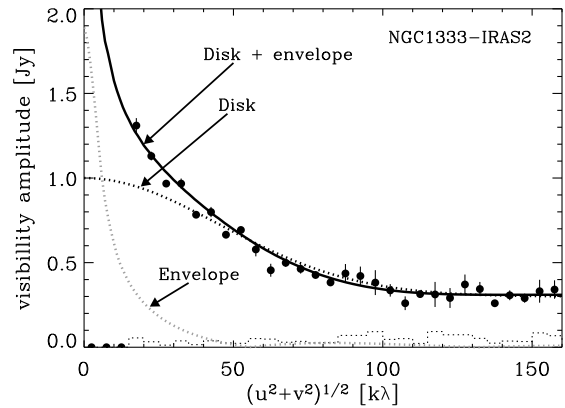}}
\caption{Interferometric 850~$\mu$m continuum observations of the
  NGC~1333-IRAS2A Class 0 YSO from the SMA \citep{iras2sma}. The data
  show the presence of a compact source of emission not accounted for
  by the larger scale envelope models. Rather, the data can be
  well-fit by a combination of the extended envelope and a 300~AU
  circumstellar disk.}\label{n1333i2_disk}
\end{figure}

\begin{figure}
\resizebox{\hsize}{!}{\rotatebox{270}{\includegraphics{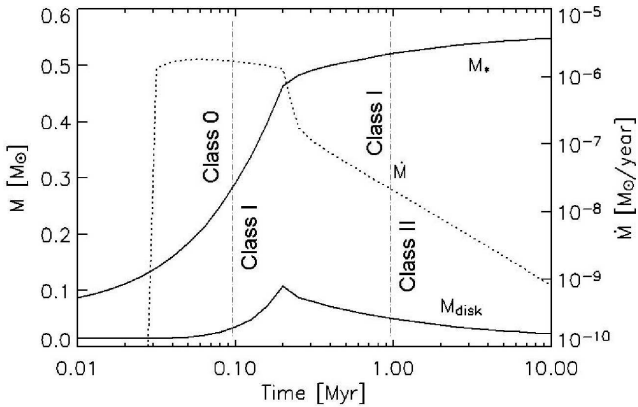}}}
\caption{Model for the evolution of disk and stellar masses (solid
  lines) and accretion rate from the disk onto the star from the Class
  0 through Class II stages of young stellar objects. Figure from
  \cite{dullemondppv} after \cite{hueso05}.}\label{huesofig}
\end{figure}

The study of disk formation in the earliest stages will be a central
scientific goal of ALMA. Key questions include whether most of the
disk mass is already assembled in the earliest Class 0 phase (e.g.,
Fig.~\ref{n1333i2_disk}), or whether disk growth --and thus stellar
growth-- continues in the later stages. Also, the dynamics of disk
formation and how this depends on initial core parameters (e.g., core
rotation) are unclear.  Through high spatial resolution kinematic
data, ALMA can image the Keplerian motions of the gas and thus obtain
direct estimates of stellar mass.  Tracing $M_{\rm env}/M_{\rm disk}$
and $M_{\rm disk}/M_*$ for a wide variety of stellar types as a
function of evolution and testing scenarios such as that shown in
Fig.~\ref{huesofig} will be a major legacy of ALMA. Multi-line
observations can determine the gas temperature and the amount of
heating through accretion shocks vs.\ UV radiation.  Finally, images
of young disks can reveal asymmetries or spiral arms
\citep[e.g.,][]{lin06}, indicative of gravitational instabilities
which may lead to giant planet formation \citep{boss03}.

\subsection{Disks around pre-main sequence stars}

{\it Q14. What is the physical and chemical structure of the
inner planet-forming zones of disks?}

Large gas disks with masses of $\sim 10^{-2}$ M$_{\odot}$ have been
revealed around classical T~Tauri and Herbig Ae stars with ages of a
few Myr by millimeter continuum and CO line emission, providing
constraints on masses and sizes of disks, their velocity patterns and
even the level of turbulence
\citep[e.g.,][]{koerner95,dutrey96,andrews07,qi03}.  However, these
data only probe the outer ($>50$AU) disk region.  ALMA will have the
sensitivity to detect and image all the dust in the disk down to the 1
AU scale.  Through multifrequency observations, ALMA will also be able
to measure the change of dust properties within disks, perhaps showing
direct evidence for radius-dependent grain growth in the midplane, up
to sizes of several mm.  ALMA will have the sensitivity to map
optically thick lines at a few AU resolution, providing information
about the gas content and its chemistry and kinematics down to the
planet-forming zones.  The stellar masses inferred from kinematics
form important direct tests of pre-main-sequence stellar evolution
models. By imaging lines with different excitation conditions, maps of
the H$_2$ density distribution will become possible.  Together, such
ALMA data can provide unbiased surveys of disks in different
star-forming regions, down to an equivalent sensitivity of a few Earth
masses of dust and gas, and probe the distribution of disk parameters
with stellar mass, luminosity, age and environment.

\subsection{Disk evolution and gap formation}

{\it Q15: When and how do gas and dust disappear from the disks?  Do
  they disappear at the same time? Are there multiple paths from
  gas-rich disks to the debris disk stage?}

Near-infrared surveys have shown that inner dust disks ($<$few AU)
disappear on timescales of a few Myr \citep{haisch01}. Surveys with
{\it Spitzer} at mid-infrared wavelengths are starting to reveal a
similar trend for the planet-forming zones out to $\sim 10$~AU, with a
significantly lower disk fraction for weak-line T Tauri stars than for
their classical counterparts \citep{cieza07}. Indeed, examination of
hundreds of SEDs of stars with disks in {\it Spitzer} surveys show
that there may be multiple evolutionary paths from the massive
gas-rich disks to the tenuous gas-poor debris disks, involving both
grain growth and gap opening. ALMA will be critical to study these
transitional objects by imaging the holes or gaps in their dust disks
down to a few AU and measure the remaining gas mass through tracers
like CO and [C~I], some of which may be left inside the holes.  Giant
planet formation, grain growth and photoevaporation are the three
major contending theories for explaining holes in the dust disks but
they have different predictions for the gas vs.\ large dust
distribution.  Overall, surveys for gaps in disks can provide
statistics on the frequency and timescale for planet formation.

\section{Conclusions}

ALMA will be vital and unique to answer key questions in star- and
planet formation, by resolving the physical processes taking place
during the collapse of molecular clouds, imaging the structure of
protostars and of proto-planetary disks, and determining the chemical
composition of the material from which future solar systems are made.
Many of the ALMA source lists will come from unbiased surveys being
carried out now, in particular {\it Spitzer}, {\it Herschel},
near-infrared and single-dish submillimeter surveys.  To extract
information from ALMA data, however, sophisticated analysis and
modeling tools are needed. The community needs to invest now in those
tools to ensure that they are ready by the time that ALMA is fully
commissioned.

Other major facilities in the timeframe of ALMA operations include the
{\it James Webb Space Telescope} and ground-based extremely large
optical telescopes (ELTs). These facilities will be highly
complementary to ALMA, each addressing a different part of the star-
and planet formation puzzle.  There is no doubt, however, that ALMA
will be {\it the} key instrument for much of the physics and chemistry
associated with star- and planet formation.


\begin{thebibliography}{60}
\expandafter\ifx\csname natexlab\endcsname\relax\def\natexlab#1{#1}\fi

\bibitem[{{Allen} {et~al.}(2007){Allen}, {Megeath}, {Gutermuth}, {Myers},
  {Wolk}, {Adams}, {Muzerolle}, {Young}, \& {Pipher}}]{allenppv}
{Allen}, L., {Megeath}, S.~T., {Gutermuth}, R., {et~al.} 2007, in Protostars
  and Planets V, ed. B.~{Reipurth}, D.~{Jewitt}, \& K.~{Keil}, 361--376

\bibitem[{{Alves} {et~al.}(2007){Alves}, {Lombardi}, \& {Lada}}]{alves07}
{Alves}, J., {Lombardi}, M. \& {Lada}, C.~J. 2007, \aap, 462, L17


\bibitem[{{Andrews} \& {Williams}(2007)}]{andrews07}
{Andrews}, S.~M. \& {Williams}, J.~P. 2007, \apj, 659, 705

\bibitem[{{Arce} \& {Sargent}({2006})}]{arce06}
{Arce}, H.~G. \& {Sargent}, A.~I. {2006}, \apj, 646, 1070

\bibitem[{{Bachiller} \& {Tafalla}(1999)}]{bachillercreteII}
{Bachiller}, R. \& {Tafalla}, M. 1999, in The Origin of Stars and Planetary
  Systems. ed C.\ J. Lada \& N.\ D. Kylafis. (Kluwer Academic Publishers,
  Dordrecht), 227

\bibitem[{{Bate} \& {Bonnell}(2005)}]{bate05}
{Bate}, M.~R. \& {Bonnell}, I.~A. 2005, \mnras, 356, 1201

\bibitem[{{Bergin} {et~al.}(2007){Bergin}, {Aikawa}, {Blake}, \& {van
  Dishoeck}}]{berginppv}
{Bergin}, E.~A., {Aikawa}, Y., {Blake}, G.~A., \& {van Dishoeck}, E.~F. 2007,
  in Protostars and Planets V, ed. B.~{Reipurth}, D.~{Jewitt}, \& K.~{Keil},
  751--766

\bibitem[{{Bergin} {et~al.}(2002){Bergin}, {Alves}, {Huard}, \&
  {Lada}}]{bergin02}
{Bergin}, E.~A., {Alves}, J., {Huard}, T., \& {Lada}, C.~J. 2002, \apjl, 570,
  L101

\bibitem[{{Bisschop} {et~al.}(2007){Bisschop}, {J{\o}rgensen}, {van Dishoeck},
  \& {de Wachter}}]{bisschop07}
{Bisschop}, S.~E., {J{\o}rgensen}, J.~K., {van Dishoeck}, E.~F., \& {de
  Wachter}, E.~B.~M. 2007, \aap, 465, 913

\bibitem[{{Bonnell} \& {Bate}(2006)}]{bonnell06}
{Bonnell}, I.~A. \& {Bate}, M.~R. 2006, \mnras, 370, 488

\bibitem[{{Boss}(2003)}]{boss03}
{Boss}, A.~P. 2003, \apj, 599, 577

\bibitem[{{Bottinelli} {et~al.}(2007){Bottinelli}, {Ceccarelli}, {Williams}, \&
  {Lefloch}}]{bottinelli07}
{Bottinelli}, S., {Ceccarelli}, C., {Williams}, J.~P., \& {Lefloch}, B. 2007,
  \aap, 463, 601

\bibitem[{{Bourke} {et~al.}(2005){Bourke}, {Crapsi}, {Myers}, {Evans},
  {Wilner}, {Huard}, {J{\o}rgensen}, \& {Young}}]{bourke05}
{Bourke}, T.~L., {Crapsi}, A., {Myers}, P.~C., {et~al.} 2005, \apjl, 633, L129

\bibitem[{{Bourke} {et~al.}({2006}){Bourke}, {Author}, {Author}, {Author},
  {Author}, {Author}, {Author}, {Author}, {Author}, {Author}, \&
  {Author}}]{bourke06}
{Bourke}, T.~L., {Myers}, P.~C., {Evans}, N.~J., {et~al.} {2006}, \apjl, 649,
L37

\bibitem[{{Caselli} {et~al.}(1999){Caselli}, {Walmsley}, {Tafalla}, {Dore}, \&
  {Myers}}]{caselli99}
{Caselli}, P., {Walmsley}, C.~M., {Tafalla}, M., {Dore}, L., \& {Myers}, P.~C.
  1999, \apjl, 523, L165

\bibitem[{{Ceccarelli} {et~al.}(2007){Ceccarelli}, {Caselli}, {Herbst},
  {Tielens}, \& {Caux}}]{ceccarellippv}
{Ceccarelli}, C., {Caselli}, P., {Herbst}, E., {Tielens}, A.~G.~G.~M., \&
  {Caux}, E. 2007, in Protostars and Planets V, ed. B.~{Reipurth} et al.,
47--62

\bibitem[{{Cesaroni}(2005)}]{cesaroni05review}
{Cesaroni}, R. 2005, \apss, 295, 5

\bibitem[{{Cieza} {et~al.}({2007}){Cieza}, {Padgett}, {Stapelfeldt},
  {Augereau}, {Harvey}, {Evans}, {Merin}, {Koerner}, {Sargent}, {Author}, \&
  {Author}}]{cieza07}
{Cieza}, L., {Padgett}, D.~L., {Stapelfeldt}, K.~R., {et~al.} {2007}, \apj,
  {in press}

\bibitem[{{Crapsi} {et~al.}(2005){Crapsi}, {Caselli}, {Walmsley}, {Myers},
  {Tafalla}, {Lee}, \& {Bourke}}]{crapsi05}
{Crapsi}, A., {Caselli}, P., {Walmsley}, C.~M., {et~al.} 2005, \apj, 619, 379

\bibitem[{{Di Francesco} {et~al.}(2001){Di Francesco}, {Myers}, {Wilner},
  {Ohashi}, \& {Mardones}}]{difrancesco01}
{Di Francesco}, J., {Myers}, P.~C., {Wilner}, D.~J., {Ohashi}, N., \&
  {Mardones}, D. 2001, \apj, 562, 770

\bibitem[{{Dullemond} {et~al.}(2007){Dullemond}, {Hollenbach}, {Kamp}, \&
  {D'Alessio}}]{dullemondppv}
{Dullemond}, C.~P., {Hollenbach}, D., {Kamp}, I., \& {D'Alessio}, P. 2007, in
  Protostars and Planets V, ed. B.~{Reipurth}, D.~{Jewitt}, \& K.~{Keil},
  555--572

\bibitem[{{Dunham} {et~al.}(2006){Dunham}, {Evans}, {Bourke}, {Dullemond},
  {Young}, {Brooke}, {Chapman}, {Myers}, {Porras}, {Spiesman}, {Teuben}, \&
  {Wahhaj}}]{dunham06}
{Dunham}, M.~M., {Evans}, N.~J. II, {Bourke}, T.~L., {et~al.} 2006, \apj, 651,
  945

\bibitem[{{Dutrey} {et~al.}(1996){Dutrey}, {Guilloteau}, {Duvert}, {Prato},
  {Simon}, {Schuster}, \& {Menard}}]{dutrey96}
{Dutrey}, A., {Guilloteau}, S., {Duvert}, G., {et~al.} 1996, \aap, 309, 493

\bibitem[{{Haisch} {et~al.}(2001){Haisch}, {Lada}, \& {Lada}}]{haisch01}
{Haisch}, Jr., K.~E., {Lada}, E.~A., \& {Lada}, C.~J. 2001, \apjl, 553, L153

\bibitem[{{Hogerheijde}(2001)}]{hogerheijde01}
{Hogerheijde}, M.~R. 2001, \apj, 553, 618

\bibitem[{{Hollis} {et~al.}(2000){Hollis}, {Lovas}, \& {Jewell}}]{hollis00}
{Hollis}, J.~M., {Lovas}, F.~J., \& {Jewell}, P.~R. 2000, \apjl, 540, L107

\bibitem[{{Hollis} {et~al.}(2006){Hollis}, {Lovas}, {Remijan}, {Jewell},
  {Ilyushin}, \& {Kleiner}}]{hollis06}
{Hollis}, J.~M., {Lovas}, F.~J., {Remijan}, A.~J., {et~al.} 2006, \apjl, 643,
  L25

\bibitem[{{Hueso} \& {Guillot}(2005)}]{hueso05}
{Hueso}, R. \& {Guillot}, T. 2005, \aap, 442, 703

\bibitem[{{J{\o}rgensen}(2004)}]{l483art}
{J{\o}rgensen}, J.~K. 2004, \aap, {424}, 589

\bibitem[{{J{\o}rgensen} {et~al.}(2007){J{\o}rgensen}, {Bourke}, {Myers}, {Di
  Francesco}, {van Dishoeck}, {Lee}, {Ohashi}, {Sch\"{o}ier}, {Takakuwa},
  {Wilner}, \& {Zhang}}]{prosacpaper}
{J{\o}rgensen}, J.~K., {Bourke}, T.~L., {Myers}, P.~C., {et~al.} 2007, \apj,
  659, 479

\bibitem[{{J{\o}rgensen} {et~al.}(2005){J{\o}rgensen}, {Bourke}, {Myers},
  {Sch\"{o}ier}, {van Dishoeck}, \& {Wilner}}]{iras2sma}
{J{\o}rgensen}, J.~K., {Bourke}, T.~L., {Myers}, P.~C., {et~al.} 2005, \apj,
  {632}, 973

\bibitem[{{J{\o}rgensen} {et~al.}(2006){J{\o}rgensen}, {Harvey}, {Evans},
  {Huard}, {Allen}, {Porras}, {Blake}, {Bourke}, {Chapman}, {Cieza}, {Koerner},
  {Lai}, {Mundy}, {Myers}, {Padgett}, {Rebull}, {Sargent}, {Spiesman},
  {Stapelfeldt}, {van Dishoeck}, {Wahhaj}, \& {Young}}]{perspitz}
{J{\o}rgensen}, J.~K., {Harvey}, P.~M., {Evans}, N.~J.~II., {et~al.} 2006, \apj,
  645, 1246

\bibitem[{{J{\o}rgensen} {et~al.}(2004){J{\o}rgensen}, {Sch\"{o}ier}, \& {van
  Dishoeck}}]{paperii}
{J{\o}rgensen}, J.~K., {Sch\"{o}ier}, F.~L., \& {van Dishoeck}, E.~F. 2004,
  \aap, 416, 603

\bibitem[{{Kirk} {et~al.}(2007){Kirk}, {Ward-Thompson}, \&
  {Andr{\'e}}}]{kirk07}
{Kirk}, J.~M., {Ward-Thompson}, D., \& {Andr{\'e}}, P. 2007, \mnras, 375, 843

\bibitem[{{Koerner} \& {Sargent}(1995)}]{koerner95}
{Koerner}, D.~W. \& {Sargent}, A.~I. 1995, \aj, 109, 2138

\bibitem[{{Krumholz} {et~al.}(2005){Krumholz}, {McKee}, \&
  {Klein}}]{krumholz05}
{Krumholz}, M.~R., {McKee}, C.~F., \& {Klein}, R.~I. 2005, \nat, 438, 332

\bibitem[{{Lada}(1999)}]{Lada99}
{Lada}, C.~J. 1999, in The Origin of Stars and Planetary Systems. Edited by C.\
  J. Lada \& N.\ D. Kylafis. (Kluwer Academic Publishers, Dordrecht), 143


\bibitem[{{Lahuis} {et~al.}(2006){Lahuis}, {van Dishoeck}, {Boogert},
  {Pontoppidan}, {Blake}, {Dullemond}, {Evans}, {Hogerheijde}, {J{\o}rgensen},
  {Kessler-Silacci}, \& {Knez}}]{lahuis06}
{Lahuis}, F., {van Dishoeck}, E.~F., {Boogert}, A.~C.~A., {et~al.} 2006, \apjl,
  636, L145

\bibitem[{{Lin} {et~al.}(2006){Lin}, {Ohashi}, {Lim}, {Ho}, {Fukagawa}, \&
  {Tamura}}]{lin06}
{Lin}, S.-Y., {Ohashi}, N., {Lim}, J., {et~al.} 2006, \apj, 645, 1297

\bibitem[{{Luhman} {et~al.}(2007){Luhman}, {Joergens}, {Lada}, {Muzerolle},
  {Pascucci}, \& {White}}]{luhmanppv}
{Luhman}, K.~L., {Joergens}, V., {Lada}, C., {et~al.} 2007, in Protostars and
  Planets V, ed. B.~{Reipurth}, D.~{Jewitt}, \& K.~{Keil}, 443--457

\bibitem[{{McKee} \& {Tan}(2003)}]{mckee03}
{McKee}, C.~F. \& {Tan}, J.~C. 2003, \apj, 585, 850

\bibitem[{{Mohanty} {et~al.}(2004){Mohanty}, {Jayawardhana}, {Natta},
  {Fujiyoshi}, {Tamura}, \& {Barrado y Navascu{\'e}s}}]{mohanty04}
{Mohanty}, S., {Jayawardhana}, R., {Natta}, A., {et~al.} 2004, \apjl, 609, L33

\bibitem[{{Neufeld} \& {Hollenbach}(1994)}]{neufeld94}
{Neufeld}, D.~A. \& {Hollenbach}, D.~J. 1994, \apj, 428, 170

\bibitem[{{Pi{\'e}tu} {et~al.}(2007){Pi{\'e}tu}, {Dutrey}, \&
  {Guilloteau}}]{pietu07}
{Pi{\'e}tu}, V., {Dutrey}, A., \& {Guilloteau}, S. 2007, \aap, 467, 163

\bibitem[{{Qi} {et~al.}(2003){Qi}, {Kessler}, {Koerner}, {Sargent}, \&
  {Blake}}]{qi03}
{Qi}, C., {Kessler}, J.~E., {Koerner}, D.~W., {Sargent}, A.~I., \& {Blake},
  G.~A. 2003, \apj, 597, 986

\bibitem[{{Rebull} {et~al.}(2007){Rebull}, {Stapelfeldt}, {Evans},
  {J{\o}rgensen}, {Harvey}, {Brooke}, {Bourke}, {Padgett}, {Chapman}, {Lai},
  {Spiesmann}, {Noreiga-Crespo}, {Merin}, {Huard}, {Allen}, {Blake}, {Jarrett},
  {Koerner}, {Mundy}, {Myers}, {Sargent}, {van Dishoeck}, {Wahhaj}, \&
  {Young}}]{rebull07}
{Rebull}, L.~M., {Stapelfeldt}, K.~R., {Evans}, N.~J.~II, {et~al.} 2007, ApJS,
171, 447

\bibitem[{{Richer} {et~al.}(2000){Richer}, {Shepherd}, {Cabrit}, {Bachiller},
  \& {Churchwell}}]{richerppiv}
{Richer}, J.~S., {Shepherd}, D.~S., {Cabrit}, S., {Bachiller}, R., \&
  {Churchwell}, E. 2000, in Protostars and Planets IV, ed. V. Mannings, A.\ P.
  Boss, \& S.\ S. Russell, 867

\bibitem[{{Shang} {et~al.}(2007){Shang}, {Li}, \& {Hirano}}]{shangppv}
{Shang}, H., {Li}, Z.-Y., \& {Hirano}, N. 2007, in Protostars and Planets V,
  ed. B.~{Reipurth}, D.~{Jewitt}, \& K.~{Keil}, 261--276

\bibitem[{{Shepherd} \& {Churchwell}(1996)}]{shepherd96}
{Shepherd}, D.~S. \& {Churchwell}, E. 1996, \apj, 457, 267

\bibitem[{{Shepherd} \& {Kurtz}(1999)}]{shepherd99}
{Shepherd}, D.~S. \& {Kurtz}, S.~E. 1999, \apj, 523, 690

\bibitem[{{Simon} {et~al.}(2006){Simon}, {Jackson}, {Rathborne}, \&
  {Chambers}}]{simon06}
{Simon}, R., {Jackson}, J.~M., {Rathborne}, J.~M., \& {Chambers}, E.~T. 2006,
  \apj, 639, 227

\bibitem[{{Tafalla} {et~al.}(1998){Tafalla}, {Mardones}, {Myers}, {Caselli},
  {Bachiller}, \& {Benson}}]{tafalla98}
{Tafalla}, M., {Mardones}, D., {Myers}, P.~C., {et~al.} 1998, \apj, 504, 900

\bibitem[{{Takahashi} {et~al.}(2006){Takahashi}, {Saito}, {Takakuwa}, \&
  {Kawabe}}]{takahashi06}
{Takahashi}, S., {Saito}, M., {Takakuwa}, S., \& {Kawabe}, R. 2006, \apj, 651,
  933

\bibitem[{{Testi} {et~al.}(2002){Testi}, {Natta}, {Oliva}, {D'Antona},
  {Comeron}, {Baffa}, {Comoretto}, \& {Gennari}}]{testi02}
{Testi}, L., {Natta}, A., {Oliva}, E., {et~al.} 2002, \apjl, 571, L155


\bibitem[{{van der Tak} {et~al.}(2005){van der Tak}, {Caselli}, \&
  {Ceccarelli}}]{vandertak05}
{van der Tak}, F.~F.~S., {Caselli}, P., \& {Ceccarelli}, C. 2005, \aap, 439,
  195

\bibitem[{{van Dishoeck}(2006)}]{vandishoeckpnas}
{van Dishoeck}, E.~F. 2006, Proceedings of the National Academy of Science,
  103, 12249

\bibitem[{{Williams} {et~al.}(2000){Williams}, {Blitz}, \&
  {McKee}}]{williamsppiv}
{Williams}, J.~P., {Blitz}, L., \& {McKee}, C.~F. 2000, Protostars and Planets
  IV, 97

\bibitem[{{Young} {et~al.}(2004){Young}, {J{\o}rgensen}, {Shirley},
  {Kauffmann}, {Huard}, {Lai}, {Lee}, {Crapsi}, {Bourke}, {Dullemond},
  {Brooke}, {Porras}, {Spiesman}, {Allen}, {Blake}, {Evans}, {Harvey},
  {Koerner}, {Mundy}, {Myers}, {Padgett}, {Sargent}, {Stapelfeldt}, {van
  Dishoeck}, {Bertoldi}, {Chapman}, {Cieza}, {DeVries}, {Ridge}, \&
  {Wahhaj}}]{young04}
{Young}, C.~H., {J{\o}rgensen}, J.~K., {Shirley}, Y.~L., {et~al.} 2004, \apjs,
  154, 396

\end{thebibliography}

%
%

\end{document}